# A Novel Quasi-One-Dimensional Topological Insulator in Bismuth Iodide $\beta$-Bi$_4$I$_4$

Gabriel Autès,[1,2]* Anna Isaeva,[3]* Luca Moreschini,[4]* Jens C. Johannsen,[5]* Andrea Pisoni,[5] Ryo Mori,[6,7] Wentao Zhang,[6,8] Taisia G. Filatova,[9] Alexey N. Kuznetsov,[9] László Forró,[5] Wouter Van den Broek,[10] Yeongkwan Kim,[4,11] Keun Su Kim,[12,13] Alessandra Lanzara,[6,8] Jonathan D. Denlinger,[4] Eli Rotenberg,[4] Aaron Bostwick,[4] Marco Grioni[5] & Oleg V. Yazyev[1,2]†

[1] Institute of Theoretical Physics, Ecole Polytechnique Fédérale de Lausanne (EPFL), CH-1015 Lausanne, Switzerland

[2] National Center for Computational Design and Discovery of Novel Materials MARVEL, Ecole Polytechnique Fédérale de Lausanne (EPFL), CH-1015 Lausanne, Switzerland

[3] Department of Chemistry and Food Chemistry, TU Dresden, D-01062 Dresden, Germany

[4] Advanced Light Source (ALS), Lawrence Berkeley National Laboratory, Berkeley, California 94720, USA

[5] Institute of Condensed Matter Physics, Ecole Polytechnique Fédérale de Lausanne (EPFL), CH-1015 Lausanne, Switzerland

[6] Materials Sciences Division, Lawrence Berkeley National Laboratory, Berkeley, California 94720, USA

[7] Graduate Group in Applied Science and Technology, University of California, Berkeley, California 94720, USA

[8] Department of Physics, University of California, Berkeley, CA 94720, USA

[9] Department of Chemistry, Lomonosov Moscow State University, Leninskie Gory 1-3, GSP-1, 119991 Moscow, Russian Federation

[10] Experimental Physics, Ulm University, Albert-Einstein-Allee 11, D-89081 Ulm, Germany

[11] Institute of Physics and Applied Physics, Yonsei University, Seoul 120-749, Korea

[12] Departement of Physics, Pohang University of Science and Technology, Pohang 790-784, Korea

[13] Center for Artificial Low Dimensional Electronic Systems, Institute for Basic Science, Pohang 790-784, Korea

* These authors contributed equally to this work.
† Corresponding author. E-mail: oleg.yazyev@epfl.ch

**Recent progress in the field of topological states of matter[1,2] has largely been initiated by the discovery of bismuth and antimony chalcogenide bulk topological insulators (TIs)[3–6], followed by closely related ternary compounds[7–16] and predictions of several weak TIs[17–19]. However, both the conceptual richness of $Z_2$ classification of TIs as well as their structural and compositional diversity are far from being fully exploited. Here, a new $Z_2$ topological insulator is theoretically predicted and experimentally confirmed in the $\beta$-phase of quasi-one-dimensional bismuth iodide $Bi_4I_4$. The electronic structure of $\beta$-$Bi_4I_4$, characterized by $Z_2$ invariants (1;110), is in proximity of both the weak TI phase (0;001) and the trivial insulator phase (0;000). Our angle-resolved photoemission spectroscopy measurements on the (001) surface reveal a highly anisotropic band-crossing feature located at the $\overline{M}$ point of the surface Brillouin zone and showing no dispersion with the photon energy, thus being fully consistent with the theoretical prediction.**

The $\beta$-phase of the quasi-one-dimensional (quasi-1D) bismuth iodide $Bi_4I_4$ crystallizes in a monoclinic base-centered lattice (space group $C12/m1$ (No. 12), $a$ = 14.386 Å, $b$ = 4.430 Å, $c$ = 10.493 Å and $\beta$ = 107.9°), the structure is shown in Fig. 1a[20]. $\beta$-$Bi_4I_4$ belongs to a family of bismuth-rich iodides, which also includes $\alpha$-$Bi_4I_4$, $Bi_{14}I_4$, $Bi_{16}I_4$ and $Bi_{18}I_4$, all composed of one-dimensional molecular fragments held together by weaker non-covalent interactions[21]. Two modifications of the $Bi_4I_4$ crystal structure, $\alpha$- and $\beta$-, arise from the differing stacking of molecular chains, whose structure and composition remain unchanged. The 1D building blocks of $\beta$-$Bi_4I_4$, aligned along the $b$ axis, can be viewed as narrow nanoribbons of bismuth bilayer (4 Bi atoms in width) terminated by iodine atoms (Fig. 1b). The crystal structure of $\beta$-$Bi_4I_4$ features two types of non-equivalent bismuth atoms: peripheral $Bi_A$ atoms saturated by covalent bonds to 4 iodine atoms and inner $Bi_B$ atoms which bind only to 3 bismuth atoms.

Previously reported crystal-growth techniques [20,22] were optimized to produce $\beta$-$Bi_4I_4$ single crystals up to 10 × 1 × 0.5 mm in size (Fig. 1c and Methods). The crystals demonstrate a high degree of crystalline order with no intergrown domains, stacking faults or other defects (Supplementary Information). The material thus appears to be practically defect-free, which implies a low concentration of intrinsic charge carriers. The anisotropy of the $\beta$-$Bi_4I_4$ crystal structure is reflected in the needle-like shape of the single crystals. The longest dimension corresponds to the $b$ axis, i.e. oriented along the one-dimensional chains. Closer inspection of the crystals reveals the presence of cleavage planes parallel to that direction (Fig. 1d).

The motivation for the experimental investigation was the result of first-principles electronic structure calculations showing that $\beta$-$Bi_4I_4$ would exhibit a topologically non-trivial phase (see Methods and Supplementary Information). The electronic band structure of $\beta$-$Bi_4I_4$ calculated using the density functional theory (DFT) approach, along a path in $k$-space including all inequivalent time-reversal invariant momentum (TRIM) points in the bulk BZ (indicated in Fig. 2a), is shown in Fig. 2b (dashed lines). These calculations predict an indirect band gap of 0.158 eV, with a valence band (VB) maximum at the $\Gamma$ point and a conduction band (CB) minimum at the M point. The band dispersion is relatively weak along the A$\Gamma$YM path, which corresponds to the inter-chain bonding directions, and strong along the BΓ direction oriented parallel to the chains, thus reflecting the quasi-1D character of $\beta$-Bi$_4$I$_4$. This compound has a centrosymmetric crystal structure, hence its $Z_2$ topological invariants ($v_0;v_1v_2v_3$) can be computed using the method of Fu and Kane[23]. According to our DFT calculations, $\beta$-Bi$_4$I$_4$ would be a weak TI with the $Z_2$ indices (0;001). The topological non-triviality of this material stems from the spin-orbit-induced band inversion at the M and Y TRIM points (Fig. 2c). The contributing electronic states are bismuth $p$ orbitals of even parity, localized on Bi$_A$ atoms (Bi$_A$ $p^+$), and of odd parity, localized on Bi$_B$ atoms (Bi$_B$ $p^-$), as was also shown for Bi$_4$Br$_4$ which has isostructural quasi-1D building blocks[24]. However, according to DFT calculations of Ref. 24 bulk Bi$_4$Br$_4$ is a trivial insulator owing to a different packing of chains.

We note that the small energy difference of 0.163 eV between VB and CB states at the M point casts doubt on the results of the DFT calculations, which are known to have limited predictive power when it comes to materials in the proximity of topological phase transitions[25]. The inaccuracy of the standard DFT method can be corrected using rigorous first-principles many-body perturbation theory approaches, such as the *GW* approximation[26,27]. This methodology has proven to provide a quantitative description of the electronic structure of bismuth chalcogenide TIs[28–30]. Indeed, our *GW* calculations show that the band inversion takes place only at the Y TRIM point (Fig. 2c). This changes the $Z_2$ indices to (1;110), thus classifying $\beta$-Bi$_4$I$_4$ as a strong TI, with no other known or proposed materials in this topological class. The change of topological class has a pronounced effect on the band dispersion of topologically protected surface states. For the DFT-predicted (0;001) weak TI phase, there are no topological surface states present at the (001) surface that is parallel to the $b$ direction of the crystal. In contrast, the (100) surface, which is also oriented along the $b$ direction, hosts topological states with a characteristic Dirac "groove" dispersion (schematically shown Fig. 2d). These surface states show linear dispersion along $y$, but are practically dispersionless along $z$ (Supplementary Figure S6). In contrast, the strong topological protection in the (1;110) $Z_2$ class materials implies the presence of topologically protected states characterized by a Dirac cone dispersion at all surfaces, independent of their orientation. Figs. 2e and 2f show the surface-projected bulk band structure and momentum-resolved local density of states at the (001) surface of $\beta$-Bi$_4$I$_4$, obtained from our calculations employing *GW* corrections. Both bulk and surface states are revealed in these plots. Magnified views of the band dispersion, in which valence and conduction bands are distinguished from the Dirac fermion surface states filling the small band gap (direct band gap of $E_g = 0.037$ eV at the Y point, according to our *GW* calculations), are shown in the insets of Figs. 2e and 2f. The Dirac point of the surface states is located at the $\overline{\text{M}}$ point of the surface BZ, which corresponds to the projection of the M and Y points of the bulk BZ (Fig. 2a). This is in striking contrast to the binary and ternary bismuth-based layered TIs of the (1;000) class, in which the Dirac cone of the surface states is located at the $\overline{\Gamma}$ point. Another important difference is related to the quasi-1D structure of $\beta$-Bi$_4$I$_4$, which results in a strong anisotropy of the surface-state Dirac fermions. According to the *GW* calculations, the Fermi velocity of the Dirac fermion surface states along the chain-like building block ($v_{Fy} = 0.57 \times 10^6$ m/s) is over five times larger than that in the perpendicular direction ($v_{Fx} = 0.10 \times 10^6$ m/s).

In order to further elaborate on the topological classification of $\beta$-Bi$_4$I$_4$, we discuss the topological phase diagram introduced in Fig. 2g. In this phase diagram, the role of intensive parameter is played by the energy-independent quasiparticle shift, $\Delta E_{QP}$, which is a shift of the conduction bands with respect to the valence bands before spin-orbit interaction is introduced[28]. The origin $\Delta E_{QP} = 0$ eV corresponds to the results of DFT calculations predicting (0;001) $Z_2$ indices for this material. Upon increasing $\Delta E_{QP}$, the system initially undergoes a topological phase transition to the (1;110) phase ($\Delta E_{QP} = 0.22$ eV), followed by another transition to the (0;000) trivial insulator phase ($\Delta E_{QP} = 0.50$ eV). The latter phase is characterized by the absence of band inversion. As expected, both topological phase transitions are characterized by the complete closure of the band gap. The region of existence of the (1;110) phase is rather narrow, which suggests that $\beta$-Bi$_4$I$_4$ can be easily shifted across both topological phase transitions, for example by means of applied external pressure or by tuning the chemical composition. The results of aforementioned $GW$ calculations roughly correspond to $\Delta E_{QP} = 0.428$ eV, which places $\beta$-Bi$_4$I$_4$ in the middle of the region of existence of the (1;110) phase. These calculations are supported by resistivity measurements that confirm a small band gap $E_g = 0.036$ eV in this material (Supplementary Information). A more comprehensive experimental evidence supporting our theoretical findings comes from angle-resolved photoemission spectroscopy (ARPES), which allows a mapping of the band structure over the whole reciprocal space.

Our ARPES measurements were performed on *in situ* cleaved single crystals of $\beta$-Bi$_4$I$_4$ that reproducibly exposed the (001) surface (see Methods). This is confirmed by the periodicity of a typical ARPES intensity map measured at $E_{bind} = 0.040$ eV, shown in Fig. 3a. Figure 3b shows a broad energy and momentum scan along the high-symmetry direction $\bar{\Gamma}' - \bar{M} - \bar{\Gamma}'$ ($k_x = 0.45$ Å$^{-1}$), which follows the boundary of the first surface BZ and is parallel to the chain direction. A prominent $\Lambda$-shaped feature representing a crossing of nearly linearly dispersing bands can be noted at the $\bar{M}$ point, alongside parabolic bands at $\bar{M}$ and $\bar{\Gamma}$ points of the surface BZ. Figs. 3c and 3d reveal the details of the electronic structure around the $\bar{M}$ point. Firstly, we note an excellent agreement with the results of our $GW$ calculations (Figs. 2e and 2f), reproducing both the $\Lambda$-shaped feature at the $\bar{M}$ point, as well as a parabolic band at the $\bar{\Gamma}$ point ($E_{bind} = 0.3$ eV) and another at the $\bar{M}$ point ($E_{bind} = 0.8$ eV). The Fermi velocities extracted from the nearly linear dispersion of the bands forming the $\Lambda$-shaped feature along the $x$ and $y$ directions in proximity of its top are $v_{Fy} = 0.60(4) \times 10^6$ m/s and $v_{Fx} = 0.1(1) \times 10^6$ m/s, respectively. These values, as well as the positions of the parabolic bands, are in excellent agreement with the results of our $GW$ calculations. The band crossing forming the $\Lambda$-shaped feature, hitherto referred to as just crossing, indicates the expected location of the small band gap of $\beta$-Bi$_4$I$_4$ that is predicted to host topologically non-trivial states. The position of the crossing observed at a binding energy of approximately $E_{bind} = 0.060$ eV suggests that $\beta$-Bi$_4$I$_4$ is an *n*-type semiconductor, in agreement with previous transport measurements[22], with a small concentration of intrinsic charge carriers compared to binary bismuth chalcogenide TIs[5,6]. The sole observation of such a prominent gapless feature rules out the other two phases, (0;001) and (0;000), for which the (001) surface is predicted to be gapped, thus proving the existence of the (1;110) strong TI phase in $\beta$-Bi$_4$I$_4$. Further investigation the $\Lambda$-shaped band using photons in a broad range of energies shows the constancy of both its dispersion and the position of the crossing point, therefore strongly supporting its surface-state origin (Supplementary Figure S9). The $k_z$ dispersion measurements also rule out the possibility that the observed crossing is of Dirac semimetal type, which could be realized exactly at a transition between two different phases present in topological phase diagram (Fig. 2g).

We note, however, that due to the small size of the bulk band gap and a very low $k_z$ dispersion of the bulk valence band (15 meV along the M−Y direction according to our *GW* calculations, cf. Fig. 2b) it is not possible to unambiguously discern bulk states from the topological surface states at the crossing in our ARPES measurements. As seen in Fig. 3e, the amplitude of the surface states largely outweighs those of the bulk states in the surface region, and therefore the former dominate the ARPES signal. This is not uncommon also in the bismuth chalcogenides TIs, where the bulk states often show much smaller ARPES intensity compared to the surface Dirac cone, but can still be detected owing to a larger magnitude of the gap.[5] Figure 3f shows a close-up of the $\Lambda$-shaped band at the $\overline{\text{M}}$ point measured with a 6 eV laser. The angle and energy resolutions are comparable to the synchrotron measurements, but due to the low photon energy the momentum resolution is greatly improved. The band dispersion is now clearly defined also for the low-energy states and presents a clear picture of the band crossing at $k_y = 0$. The importance of momentum resolution for accessing the surface-state dispersion at these conditions is further illustrated in Figure 3g, where we compare two EDCs at $\overline{\text{M}}$ measured using $h\nu$ = 6 eV and $h\nu$ = 68 eV photon energies, respectively. At $h\nu$ = 68 eV the limited momentum resolution results in two broad structures with an undefined intensity in between, while at $h\nu$ = 6 eV the intensity peaks strongly at *ca.* 60 meV binding energy, and a much smaller peak induces a shoulder (arrow in Fig. 3g) that we attribute to the bulk VB maximum.

We have also investigated the band dispersion at the $\overline{\text{M}}$ point after depositing potassium atoms that donate electrons to the conduction states thus producing a rigid downward shift of the bands by up to 0.25 eV (Supplementary Figure S11). These measurements allowed a more detailed investigation of the dispersion of the bulk conduction band. By extracting the quasiparticle energies of the valence and conduction bands where the states are expected to be of bulk character (at least *ca.* 0.1 eV away from the crossing) from fits of the momentum distribution curves, we were able to estimate a band gap value of 50 meV at $k_z$ that corresponds to $h\nu$ = 73 eV (Supplementary Figure S12).

To summarize, we reported the joint theory-experiment discovery of a novel topological insulator phase in the binary bismuth iodide $\beta$-Bi$_4$I$_4$. This material is chemically and structurally well defined and is less prone to structural defects compared to bismuth chalcogenides. Even though the bulk band gap is small, the material is characterized by a low concentration of intrinsic charge carriers, and hence, the proximity of the Fermi level to the Dirac point energy of the topological surface-state band. Importantly, the quasi-1D structure of *β*-Bi$_4$I$_4$ results in highly anisotropic surface-state Dirac fermions and suggests the possibility of combining topological order with other types of ordering characteristic to one-dimensional systems, such as the charge density wave. Moreover, being a (1;110) strong TI, this material is placed in proximity to two distinct topological phase transitions, leading to a (0;001) weak TI and a (0;000) trivial insulator, which presents further opportunities for exploring topological physics.

**Acknowledgments** We thank J. H. Dil and M. Ruck for fruitful discussions, Hyungjun Lee for discussions regarding the computational methodology, Beomyoung Kim for support during the beamtime on Merlin, M. Münch, K. Zechel and A. Weiz for assistance with synthesis and SEM/EDX measurements. We are grateful to E. Schmid for ultramicrotomy, to U. Kaiser and C. T. Koch for providing beam time for the TEM characterization. G.A. and O.V.Y. acknowledges support by the Swiss NSF (grant No. PP00P2_133552), ERC project "TopoMat" (grant No. 306504) and NCCR-MARVEL. A.I. acknowledges the Priority Program 1666 "Topological Insulators" of the Deutsche Forschungsgemeinschaft (DFG, grant No. IS 250/1-1). L.M. acknowledges support by the Swiss NSF (grant No. PA00P21-36420). The Advanced Light Source is supported by the Director, Office of Science, Office of Basic Energy Sciences, of the U.S. Department of Energy under Contract No. DE-AC02-05CH11231. W.v.d.B. acknowledges the Carl-Zeiss Foundation. Electronic structure calculations have been performed at the Swiss National Supercomputing Centre (CSCS) under project s515.

**Author contributions** O.V.Y. initiated and directed this research project; G.A. performed first-principles calculations; A.I. pointed out, synthesized and characterized the material; L.M. conducted the ARPES measurements, together with R.M. and W.Z. for the laser-based experiments; L.M., J.C.J. and M.G. analyzed the ARPES data; A.P. and L.F. carried out the transport measurements; T.G.F. and A.N.K. performed and optimized sample preparation; W.V.d.B. assisted in the TEM experiments; Y.K., K.S.K., J.D.D., A.L., E.R. and A.B. assisted in the ARPES measurements and data analysis. All authors contributed to discussions and manuscript revision.

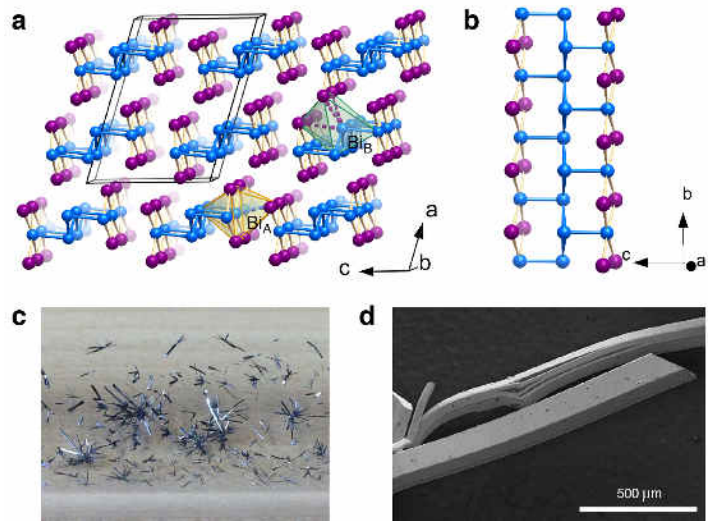

**Figure 1 | Crystal structure and single crystals of quasi-one-dimensional topological insulator *β*-Bi$_4$I$_4$. a,** Crystal structure of *β*-Bi$_4$I$_4$ viewed along the chain direction (lattice vector *b*). Bismuth and iodine atoms are shown in blue and purple, respectively. Coordination polyhedra of two non-equivalent Bi atoms, Bi$_A$ and Bi$_B$, and the conventional unit cell are indicated. Solid lines indicate covalent bonds. **b,** Atomic structure of an individual chain-like building block of *β*-Bi$_4$I$_4$. **c,** Photographic image of the crystalline product of a chemical-transport reaction on the ampoule walls. The sample is composed of *β*-Bi$_4$I$_4$ needle-like crystals, typically 2−3 mm long and few tenths of mm thick. **d,** Scanning electron microscopy (SEM) image of individual needle-like crystals of *β*-Bi$_4$I$_4$ showing the cleavage planes along the needle axis.

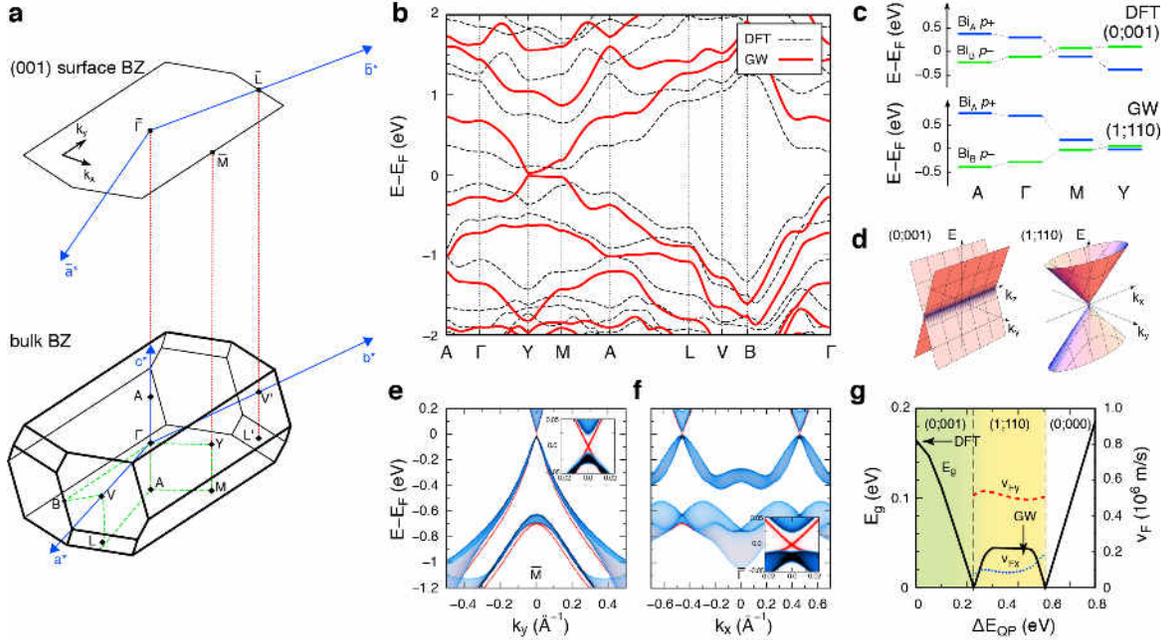

**Figure 2 | Electronic structure of *β*-Bi$_4$I$_4$ from first-principles calculations. a,** Primitive Brillouin zone (BZ) of bulk *β*-Bi$_4$I$_4$. Time-reversal invariant momentum (TRIM) points and their projection onto the (001) surface BZ are indicated. **b,** Electronic band structure along the path (green dashed line) in panel (**a**) computed using DFT and the *GW* approximation. **c,** Schematic diagram of band inversion at the TRIM points A, Γ, M and Y obtained from DFT and *GW* approximation calculations, leading to the weak (0;001) and strong (1;110) topological phases, respectively. **d,** Schematics of the topological surface-state band dispersions for weak (0;001) and strong (1;110) topological phases, respectively. **e,f,** Bulk band structure projected on the (001) surface (blue) and momentum-resolved local density of states at the (001) surface (red) along $k_y$ ($k_x$ = 0.45 Å$^{-1}$) and $k_x$ ($k_y$ = 0 Å$^{-1}$), respectively, calculated for the (1;110) phase of *β*-Bi$_4$I$_4$. The insets in panels (**e**) and (**f**) show the magnified view near the Dirac point. **g,** Topological phase diagram of *β*-Bi$_4$I$_4$ covering the (0;001) and (1;110) topological phases as well as the trivial insulator (0;000) phase. The diagram shows the magnitudes of the band gap $E_g$ and the Fermi velocities of the topological states at the (001) surface, $v_{Fx}$ and $v_{Fy}$, as a function of energy-independent quasiparticle shift $\Delta E_{QP}$. Arrows locate the DFT and *GW* results in the topological phase diagram.

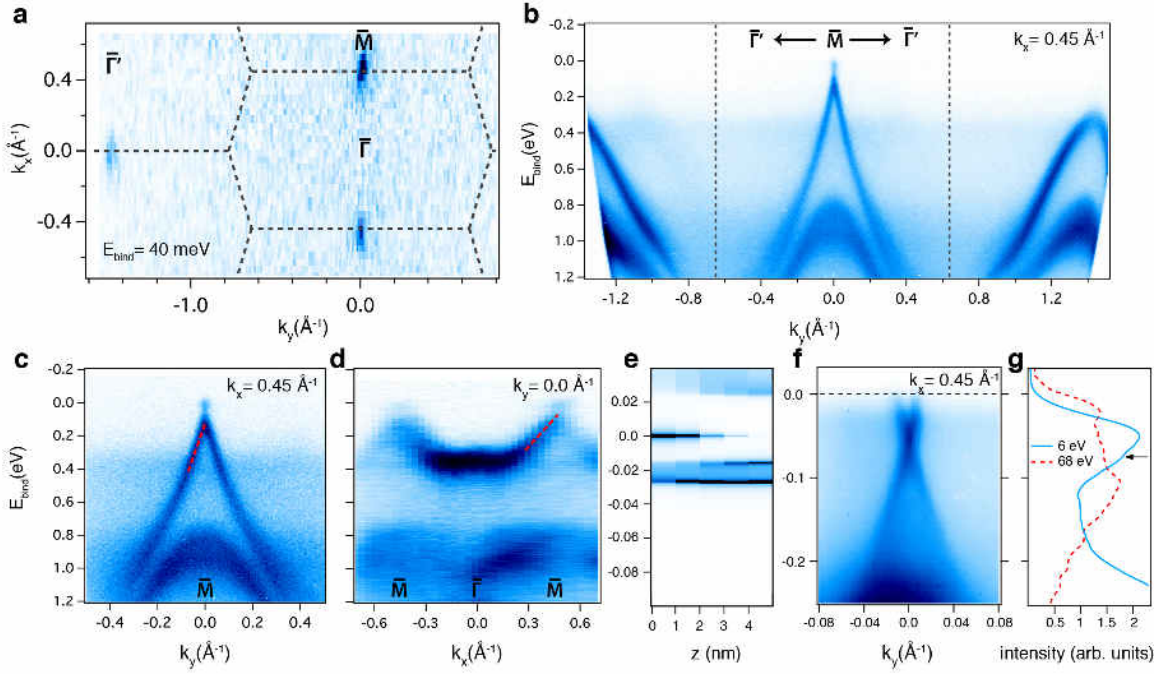

**Figure 3 | Angle-resolved photoemission spectroscopy (ARPES) spectra of *β*-Bi$_4$I$_4$. a,** Constant energy contour map at a binding energy $E_{bind}$ = 0.040 eV spanning several Brillouin zones (BZs), measured at $hv$ = 85 eV. **b,** Band dispersion covering a large energy and momentum range along the high-symmetry direction $\bar{\Gamma}' - \bar{M} - \bar{\Gamma}'$ ($k_x$ = 0.45 Å$^{-1}$) following the boundary of the first BZ. **c,** High-resolution close-up of the band dispersion around the $\bar{M}$ point measured along the *y* direction. The extracted Fermi velocity is $v_{Fy}$ = 0.60(4)×10$^6$ m/s. **d,** Band dispersion measured along $\bar{M} - \bar{\Gamma} - \bar{M}$ line ($k_y$ = 0 Å$^{-1}$) crossing the first BZ. For this direction we obtain a Fermi velocity of $v_{Fx}$ = 0.1(1)×10$^6$ m/s. **e,** Computed weights of the surface and bulk states at the $\bar{M}$ point as a function of distance to the surface and energy relative to the surface-state Dirac point. This demonstrates strong surface localization of the topological Dirac fermion states ($E$ = 0 in this plot). **f,** Details of the band dispersion around the $\bar{M}$ point measured with a UV laser ($hv$ = 6 eV). **g,** Energy dispersion curves (EDCs) at $\bar{M}$ for $hv$ = 6 eV and $hv$ = 68 eV measurements. The arrow points to the shoulder attributed to the bulk VB.

**Methods**

**Materials synthesis and characterisation.** Based on the chemical transport reactions described in Refs. 20,22, we used elemental bismuth and $HgI_2$ as starting materials in two ratios, 1:2 and 1:4, amounting to *ca*. 2 g in total. The mixtures were sealed in silica ampoules under dynamic vacuum and placed into the temperature gradient of 210/250ºC for up to 20 days. The colder end of the ampoule was tilted upwards at *ca*. 25–30 degrees. Both batches produced large crystals of *β*-$Bi_4I_4$ (up to 10 × 1 × 0.5 mm) with the needle-like morphology. It was observed that the excess of $HgI_2$ results in somewhat longer crystals, however, some of them were unusable due to surface contamination by mercury.

Crystal composition was verified by electron-dispersive X-ray spectroscopy on a Hitachi SU8020 microscope (30 kV, Oxford Silicon Drift Detector (SDD) X-Max$^N$) using Bi-*M* and I-*L* series. The crystal structure of several single crystals was tested by an X-ray diffraction study on an automated imaging plate diffractometer IPDS-II (STOE), Mo$_{K\alpha}$-radiation, room temperature. The structural refinements were in complete accordance with the earlier reported data (Supplementary Information). Furthermore, high-resolution transmission electron microscopy studies were performed on thin lamellas cut out of the single crystals by ultramicrotomy (Ultracut, Leica Microsystems) on a FEI Titan F20 $C_S$-corrected microscope (at 80 kV and at 300 kV).

**First-principles calculations.** The first-principles electronic structure calculations at the level of density functional theory (DFT) were performed employing the generalized gradient approximation (GGA)[31] as implemented in the Quantum ESPRESSO package[32]. Spin-orbit effects were accounted for using fully relativistic norm-conserving pseudopotentials acting on valence electron wavefunctions represented in the two-component spinor form[33]. The DFT-GGA calculations were carried out using an 8×8×6 *k*-point mesh and a planewave kinetic energy cutoff of 50 Ry for the wavefunctions. We used the experimentally determined crystal structure from Ref. 20.

The many-body perturbation theory calculations within the *GW* approximation were performed using the BerkeleyGW code[34]. The quasiparticle self-energy corrections were evaluated at the $G_0W_0$ level starting from non-relativistic DFT-GGA results using the approach of Hybertsen and Louie[27]. We employed a 6×6×4 *k*-point mesh, 2000 conduction bands and wavefunctions obtained using an increased planewave kinetic energy cutoff of 100 Ry for evaluating both the dielectric matrix and self-energy corrections. These parameters ensure convergence of the quasi-particle shift at the Y point with respect to the number of bands and cutoff is of the order of the meV.

The *GW*-corrected band structure and surface states were computed with the help of a tight-binding Hamiltonian constructed using the bulk Wannier functions[35] obtained from the non-relativistic DFT-GGA wavefunctions and the *GW*-corrected eigenvalues. The effects of spin-orbit interactions were included by extracting the spin-orbit Hamiltonian from relativistic DFT-GGA calculations and adding it to the *GW*-corrected non-relativistic tight-binding Hamiltonian. From the full tight-binding Hamiltonian, we computed the left and right surface Green's functions $g_L(E,k_\parallel)$ and $g_R(E,k_\parallel)$ of the semi-infinite surface models using the analytic closed form solution of Ref. 36. The corresponding bulk Green's function $G(E,k_\parallel)$ was obtained using the Dyson equation $G = \left(g_L^{-1} - t g_R t^\dagger\right)^{-1}$, where *t* is the interlayer hopping matrix. Momentum-resolved density of bulk states and momentum-resolved local density of states at the surface shown in Figs. 2e,f correspond to the imaginary part of the trace of $G$ and $g_\text{L}$, respectively.

**Photoemission measurements.** Our ARPES measurements were performed on *in situ* cleaved single crystals. The cleavage plane was consistently normal to the [001] crystallographic direction. The synchrotron data were taken at beamline 4.0.3 (Merlin) of the Advanced Light Source, Berkeley, with an angular resolution of 0.1° and a combined energy resolution better than 20 meV. We used *p* (horizontal) polarization, which provides higher flux. The laser data were measured with a ultraviolet probe laser pulse (5.93 eV) at a repetition rate of 1 MHz. The total energy resolution was ~22 meV and the angle resolution ~0.3°. Despite a value 3 times larger than on Merlin, at normal emission this translates into a momentum resolution ~0.003 Å$^{-1}$ more than twice better, thanks to the low photon energy. The beam spot size was smaller than 100 × 50 μm on Merlin and smaller than 50 × 50 μm on the laser setup. This was a key point for the outcome of the experiments given the small size of the needle-like domains in this material.

# Supplementary Information for

# A Novel Quasi-One-Dimensional Topological Insulator in Bismuth Iodide $\beta$-Bi$_4$I$_4$

Gabriel Autès,* Anna Isaeva,* Luca Moreschini,* Jens C. Johannsen,* Andrea Pisoni, Ryo Mori, Wentao Zhang, Taisia G. Filatova, Alexey N. Kuznetsov, László Forró, Wouter Van den Broek, Yeongkwan Kim, Keun Su Kim, Alessandra Lanzara, Jonathan D. Denlinger, Eli Rotenberg, Aaron Bostwick, Marco Grioni & Oleg V. Yazyev†

\* These authors contributed equally to this work.

† Corresponding author. E-mail: oleg.yazyev@epfl.ch

**CONTENTS**

**1. Sample preparation and characterization**

**2. Electronic structure calculations**

**3. Resistivity measurements**

**4. Angle-resolved photoemission spectroscopy (ARPES) measurements**

**5. Supplementary references**

# 1. Sample preparation and characterization

## 1.1. Crystal growth

In this work, a great deal of effort was paid to the optimization of the crystal growth technique for $\beta$-$Bi_4I_4$. Earlier attempts resulted in crystals of 1.5 mm in length and 0.2 mm in width, which hampered direct measurements of physical properties and made results and interpretation inconclusive[20]. A more recent contribution[22] has provided improvement of the synthetic procedure, which allowed the first experimental measurements on $\beta$-$Bi_4I_4$ crystals.

We used the same gas-phase approach as in Refs. 20,22, with bismuth metal and $HgI_2$ being the starting materials, and were able to produce crystals of 10 × 1 × 0.5 mm in size. We have optimized the synthetic protocol with respect to ampoule volume, temperature gradient and duration of annealing. Generally, longer annealing times and smaller gradients favor larger crystals; however the competing process of intergrowth is detrimental for crystal quality. It was observed that the excess of $HgI_2$ results in somewhat longer crystals, however, some of them were unusable due to a partial surface contamination by $HgI_2$.

Two Bi:$HgI_2$ ratios were used, 1:2 or 1:4, amounting to ca. 2 g in total. Silica ampoules 25 cm in length and 15 mm in diameter were dried under dynamic vacuum prior to use. Principal improvements over Ref. 22 included a smaller temperature gradient of 210/250ºC and longer experiment times, up to 20 days. The colder end of the ampoule was tilted upwards at ca. 25–30 degrees. Both batches produced crystals of $\beta$-$Bi_4I_4$ with the same needle-like morphology.

## 1.2. Sample characterization

Composition and structure of the crystal surface are decisive for subsequent ARPES studies; therefore we conducted comprehensive characterization of the obtained crystals by the following complementary methods, with the primary focus on possible non-stoichiometry and local disorder.

*Scanning Electron Microscopy* and *Electron-Dispersive X-ray Spectroscopy* were performed using a Hitachi SU8020 microscope with a triple detector system for secondary and low-energy backscattered electrons ($U_a$ = 2 kV). EDX spectra were collected using an Oxford Silicon Drift Detector (SDD) X-Max$^N$ and averaged over several dozens of crystals. Bi-*M* and I-*L* series were used for spectrum evaluation. Margins of errors are determined for each element and include the averaged error of each spectrum and the standard derivation of multiple spectra. Error propagation is considered for the molar ratio.

Average crystal composition: Bi: 52.5(1) at %; I: 47.5(5) at %.

No traces of mercury or mercury iodide were detected in samples used for further physical property measurements. A typical spectrum also contained signals of silicon, carbon and oxygen arising from glass shatters and from the sample stage. No significant variation in iodine content has been observed for the crystals.

*X-ray single-crystal diffraction:* $\alpha$- and $\beta$-modifications of $Bi_4I_4$ are reported to exist in different temperature intervals[20]; however, they only differ in the packing of molecular chains, which are flipped with respect to each other in the $\alpha$-phase, thus causing the doubling of the *c* axis. Occasional flips of the stacks or co-existing domains of two structures could hence be expected, especially given the non-equilibrium conditions of the crystal-growth (chemical transport reaction).

In order to investigate the possibility of local disorder, several single crystals were extracted from the batch and subjected to a X-ray diffraction study on an automated imaging plate diffractometer IPDS-II (STOE), Mo$K_\alpha$-radiation, λ = 71.073 pm, T = 296(1) K; numerical absorption correction[37] based on optimized crystal descriptions[38], structure solution with direct methods and refinements against $F_o^2$ [39,40]. Graphics of the structures were developed with Diamond software[41].

The present structure refinement of β-Bi$_4$I$_4$ is in full accordance with previously published data[20]. No superstructure reflections pointing towards doubling of unit cell parameters were observed (see generated reciprocal layers in Supplementary Fig. S1):

Data on structure refinement of β-Bi$_4$I$_4$: monoclinic, $C12/m1$ (No. 12), $a$ = 14.3530(6) Å, $b$ = 4.4228(2) Å, $c$ = 10.4760(4) Å, $β$ = 107.825(2)°, $V$ = 633.10(1) Å$^3$, $Z$ = 8, $r_{calc.}$ = 7.048 g cm$^{-3}$; $m$(Mo$K\alpha$) = 65.11 mm$^{-1}$; IPDS-II; $2q_{max}$ = 60°; 9862 measured, 1018 unique reflections, $R_{int}$ = 0.0182, $R_σ$ = 0.0131, 26 parameters; $R_1$[1001 $F_o$ > 4$s$($F_o$)] = 0.0132, $wR_2$(all $F_o^2$) = 0.0135, GooF = 1.071; min./max. residual electron density: −1.17 / 1.37 × 10$^{-6}$ pm$^{-3}$. Crystal size 0.033 × 0.039 × 0.266 mm

The facets of the β-Bi$_4$I$_4$ crystal were indexed (see Supplementary Fig. S2) using the X-SHAPE software[38]. The longest crystal facet runs along the direction of the crystallographic $b$ axis. This information on the crystal habit was further used for ultramicrotomy and ARPES experiments.

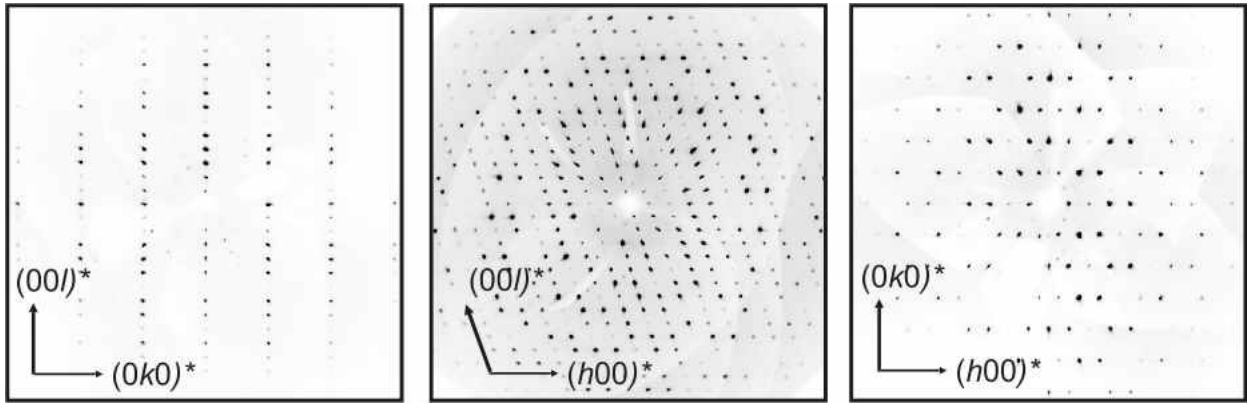

**Supplementary Figure S1.** The reciprocal [0$kl$], [$h$0$l$] and [$hk$0] planes for β-Bi$_4$I$_4$ generated from a recorded single-crystal dataset.

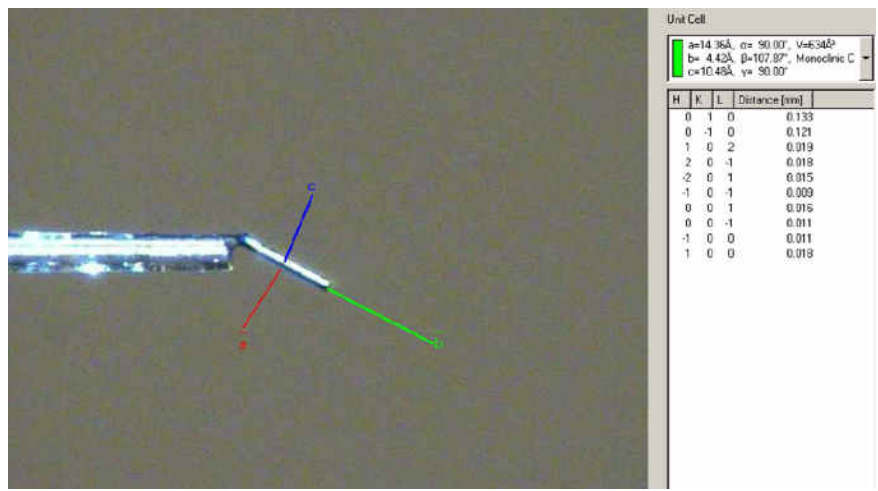

**Supplementary Figure S2.** Indexing of the *β*-Bi$_4$I$_4$ crystal facets. The footage from a video camera represents a needle-shaped crystal glued to a sample holder (on the left). The crystal dimensions along each facet are given in mm in the right panel.

*(High-Resolution) Transmission Electron Microscopy* studies were performed in order to get an insight into structural ordering and local defects on the atomic scale. The samples were prepared by cutting out thin lamellas from two *β*-Bi$_4$I$_4$ crystals perpendicular to the largest facet, i.e. along the chains, with an ultramicrotome (Ultracut, Leica Microsystems) equipped with a diamond knife. Prior to this, the crystals were embedded in the epoxy resin Epon (Fluka), that was then polymerized at 60 °C.

The nominal lamella thickness was *ca*. 50 nm, while locally crystalline flakes with a thickness of 5−15 nm were observed. High-resolution transmission electron microscopy (HRTEM) and selected area electron diffraction (SAED) studies were performed on a FEI Titan F20 microscope with C$_S$-correction operating at 80 kV and at 300 kV. No significant beam damage was observed at higher voltage. For image acquisition, a 2k × 2k Slow-Scan CCD-Camera (Gatan) was used. Image simulations were done with the JEMS software[42].

The study establishes a high degree of ordering in *β*-Bi$_4$I$_4$ crystals on the microscopic scale. Recorded images showed no stacking faults or occasional flips of the adjacent stacks of molecular chains (Supplementary Fig. S3).

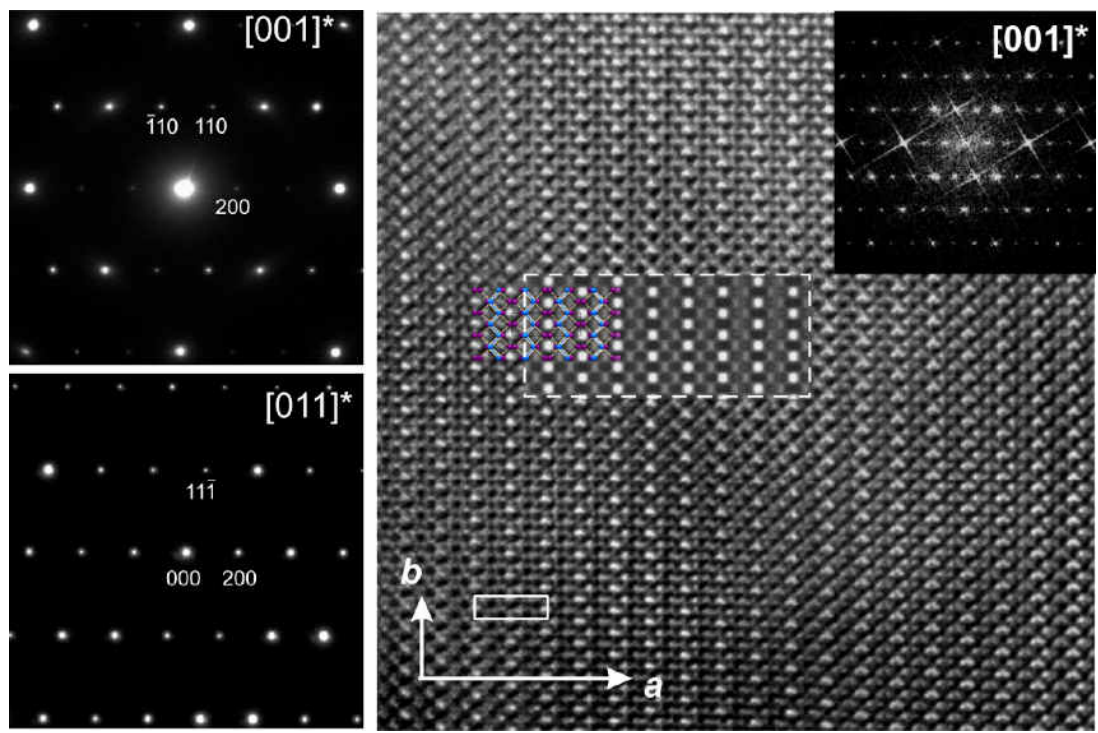

**Supplementary Figure S3.** ED (left column) and HRTEM images showing main zones of β-Bi$_4$I$_4$. The experimental ED patterns fully accord with the above given unit cell of β-Bi$_4$I$_4$ and show reflection conditions corresponding to the *C*-centered monoclinic space group. Theoretical (*f* = –90 nm, *t* = 10 nm, dashed outline) and experimental [001]*-zone HRTEM image with the Fourier transform in an inset are presented. The light dots correspond to the cavities between the adjacent molecular chains, whilst the grey dots represent voids between bismuth and iodine atoms in the chain. The unit cell is outlined.

The characterization allows us to conclude that the synthesized crystals belong solely to β-Bi$_4$I$_4$, and shows no intergrown domains, stacking faults or any other deviations from the reported crystal structure. In addition, TEM experiments demonstrate that β-Bi$_4$I$_4$ crystals can be cut by ultramicrotomy without damage or strain induction, e. g. thin lamellas (down to the thickness of 1 μm) of the compound can be prepared for further studies. Worth noting is that these samples do not exhibit prominent beam damage and do not quickly oxidize in air.

## 2. Electronic structure calculations

### *2.1. Description of computational methodology*

The first-principles electronic structure calculations at the level of density functional theory (DFT) were performed employing the generalized gradient approximation (GGA)[31] as implemented in the Quantum ESPRESSO package[32]. Spin-orbit effects were accounted for using fully relativistic norm-conserving pseudopotentials acting on valence electron wavefunctions represented in the two-component spinor form[33]. The DFT-GGA calculations were carried out using an 8×8×6 *k*-point mesh and a planewave kinetic energy cutoff of 50 Ry for the wavefunctions. We used the experimentally determined crystal structure from Ref. 20.

The many-body perturbation theory calculations within the *GW* approximation were performed using the BerkeleyGW code[34]. The quasiparticle self-energy corrections were evaluated at the $G_0W_0$ level starting from non-relativistic DFT-GGA results using the approach of Hybertsen and Louie[27]. We employed a 6×6×4 *k*-point mesh, 2000 conduction bands and wavefunctions obtained using an increased planewave kinetic energy cutoff of 100 Ry for evaluating both the dielectric matrix and self-energy corrections. These parameters ensure convergence of the quasi-particle shift at the Y point with respect to the number of bands and cutoff is of the order of the meV. Supplementary Figure S4 shows the calculated quasiparticle self-energy corrections $\Delta E_{QP}$ as a function Kohn-Sham eigenvalues $E_{DFT}$ for the valence (red) and conduction (blue) states. The calculated quasiparticle correction for the energy difference between the valence band and conduction band at the Y point is 0.428 eV (Supplementary Fig. S4). We note a relatively weak dependence of $\Delta E_{QP}$ on Kohn-Sham eigenvalues, which justifies the use of energy-independent quasiparticle shifts as an intensive parameter of the discussed topological phase diagram (Fig. 2g of the main text).

The *GW*-corrected band structure and surface states were computed with the help of a tight-binding Hamiltonian constructed using the bulk Wannier functions[35] obtained from the non-relativistic DFT-GGA wavefunctions and the *GW*-corrected eigenvalues. The effects of spin-orbit interactions were included by extracting the spin-orbit Hamiltonian from relativistic DFT-GGA calculations and adding it to the *GW*-corrected non-relativistic tight-binding Hamiltonian. From the full tight-binding Hamiltonian, we computed the left and right surface Green's functions $g_L(E,k_\parallel)$ and $g_R(E,k_\parallel)$ of the semi-infinite surface models using the analytic closed form solution of Ref. 36. The corresponding bulk Green's function $G(E,k_\parallel)$ was obtained using the Dyson equation $G = \left(g_L^{-1} - t g_R t^\dagger\right)^{-1}$, where *t* is the interlayer hopping matrix. Momentum-resolved density of bulk states and momentum-resolved local density of states at the surface shown in Figs. 2e,f of the main text correspond to the imaginary part of the trace of *G* and $g_L$ respectively.

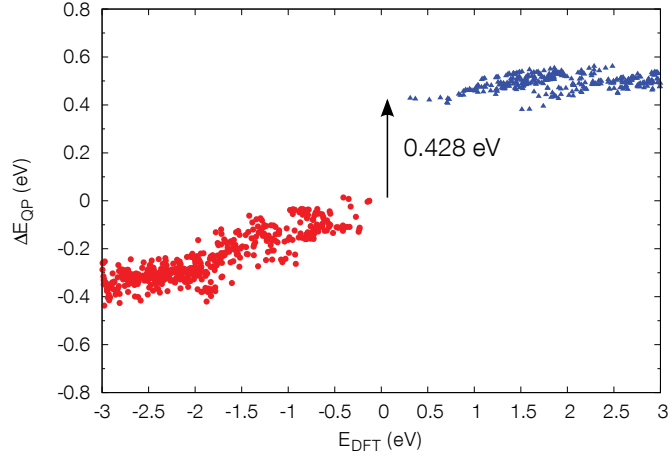

**Supplementary Figure S4.** Quasiparticle self-energy corrections $\Delta E_{QP}$ as a function of Kohn-Sham eigenvalues $E_{DFT}$ for the valence (red) and conduction (blue) states of $\beta$-Bi$_4$I$_4$.

*2.2. Discussion of $Z_2$ topological invariants*

Owing to the centrosymmetric crystal structure of $\beta$-Bi$_4$I$_4$, its $Z_2$ topological invariants $(v_0;v_1v_2v_3)$ can be evaluated following the simple recipe introduced by Fu and Kane[23]. For each time-reversal invariant momentum (TRIM) point $\Lambda$ in the primitive Brillouin zone (BZ) we compute

$$\delta_\Lambda = \prod_{i=1}^{N/2} \xi_{2i}(\Lambda),$$

where $\xi_{2i} = \xi_{2i+1}$ is the parity eigenvalue of the pair of occupied degenerate bands $2i$ and $2i+1$. The values of $\delta_\Lambda$ for the $\beta$-phase of Bi$_4$I$_4$ are reported in Supplementary Table S1. A change of sign of $\delta_\Lambda$ between two TRIM points reveals the presence of a band inversion. The strong topological index $v_0$ given by

$$(-1)^{v_0} = \prod_\Lambda \delta_\Lambda$$

is non-zero if there is an odd number of band inversions in the bulk BZ. The three weak indices $v_1$, $v_2$ and $v_3$ are obtained from the product of $\delta_\Lambda$ at 4 coplanar TRIM points in the BZ. The corresponding planes shown in Supplementary Fig. S5 give $(-1)^{v_1} = \delta_Y \delta_M \delta_V \delta_L$, $(-1)^{v_2} = \delta_Y \delta_M \delta_{V'} \delta_{L'}$ and $(-1)^{v_3} = \delta_A \delta_Y \delta_L \delta_{L'}$.

In the absence of spin-orbit interactions both DFT and the *GW* approximation calculations predict the (0;000) topologically trivial phase. The spin-orbit coupling induces a band inversion at the M and Y points resulting in a weak topological insulator phase with $Z_2$ indices (0;001) when calculations are performed at the DFT level. However, in the *GW* approximation calculations the band inversion occurs only at the Y point of the BZ. This corresponds to the (1;110) strong topological insulator phase, also confirmed by the experiments reported in this work.

**Supplementary Table S1**. Values of $\delta_\Lambda$ at the 8 TRIM points of the primitive Brillouin zone evaluated at the DFT and $GW$ level of theory, with and without spin-orbit (SO) interactions taken into account.

|  | coordinate of TRIM point | DFT and $GW$ w/o SO | DFT with SO | $GW$ with SO |
|---|---|---|---|---|
| $\delta_\Gamma$ | (0,0,0) | −1 | −1 | −1 |
| $\delta_A$ | (0,0,½) | −1 | −1 | −1 |
| $\delta_M$ | (½,½,½) | −1 | 1 | −1 |
| $\delta_Y$ | (½,½,0) | −1 | 1 | 1 |
| $\delta_V$ | (½,0,0) (0,½,0) | 1 | 1 | 1 |
| $\delta_L$ | (½,0,½) (0,½,½) | 1 | 1 | 1 |
| $(\nu_0;\nu_1\nu_2\nu_3)$ |  | (0;000) | (0;001) | (1;110) |

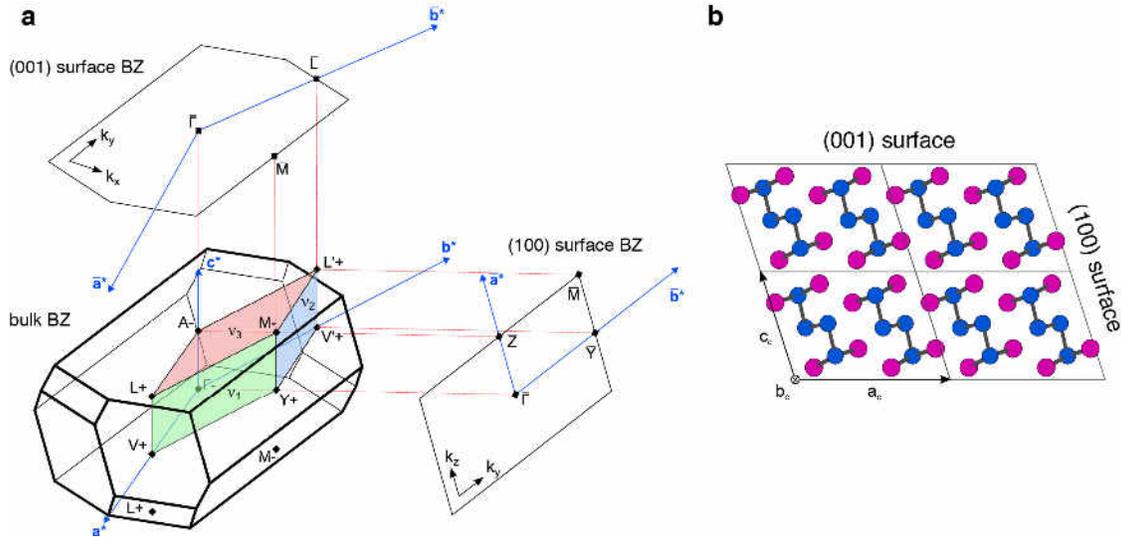

**Supplementary Figure S5. a,** Primitive Brillouin zone of $\beta$-Bi$_4$I$_4$ showing the 8 TRIM points and the 3 planes (colored) defining the weak topological indices $\nu_1$, $\nu_2$ and $\nu_3$. The signs correspond to $\delta_\Lambda$ of the strong topological phase (1;110). **b,** Crystal structure of $\beta$-Bi$_4$I$_4$ showing the (001) and (100) surfaces oriented parallel to the chain direction $b$.

### 2.3. Surface states at the (100) and (001) surfaces in the weak and strong topological phases

To complete the picture presented in the main text (Fig. 2e,f), we investigate the electronic structure of the (001) and (100) surfaces in the two discussed topologically non-trivial phases (Supplementary Fig. S6). In the (1;110) strong TI phase, the Dirac cones are present at both surfaces at the TRIM point of the surface BZ, on which the Y point of the bulk BZ is projected (*i.e.* $\bar{M}$ for the (001) surface and $\bar{\Gamma}$ for the (100) surface; cf. Supplementary Fig. S5). The dispersion is considerably stronger along the $k_y$ direction, which corresponds to the chain direction. This is a direct manifestation of the quasi-1D character of $\beta$-Bi$_4$I$_4$.

In the (0;001) weak topological phase, the topological surfaces states are not expected to be present on every surface. Using the values of $\delta_\Lambda$ at the 3D TRIM points, it is possible to predict which surface will exhibit topologically states. Following Ref. 23, for each 2D TRIM point $\Lambda$ in the 2D BZ of a given surface, we calculate

$$\pi_{\bar{\Lambda}} = \delta_{\Lambda_1} \delta_{\Lambda_2},$$

where $\Lambda_1$ and $\Lambda_2$ are the 3D TRIM points which project onto the surface BZ are $\bar{\Lambda}$. The change of sign of $\pi_{\bar{\Lambda}}$ between the 2D TRIM points $\bar{\Lambda}$ signals the existence of topological surface states. The values of $\pi_{\bar{\Lambda}}$ for the (100) and (001) surfaces calculated for the weak topological phase (0;001) are reported in Supplementary Table S2. From these values, we can conclude that in the weak topological phase the (001) surface has no topologically protected surface states, while the (100) surface has a surface state connecting the $\bar{\Gamma}\bar{Z}$ line to the $\bar{Y}\bar{M}$ line. The parity analysis is confirmed by the surface local density of states (LDOS) calculations (Supplementary Fig. S6). These calculations show that the (001) surface is gapped while the (100) surface exhibits surface states with an evident crossing along $k_y$ and a very weak dispersion along $k_z$. The existence of gapless surface states only at the (100) surface of the weak phase can also be understood in terms of the quantum spin Hall effect. A weak topological insulator with invariants (0,001) can be seen as the stacking of 2D layers of the quantum spin Hall phase in the [001] direction. These layers possess gapless edge states. Any surface formed by these edges, for instance the (100) surface, will thus exhibit highly anisotropic Dirac fermion states. On the contrary, the (001) surface that corresponds to the layer planes remains gapped.

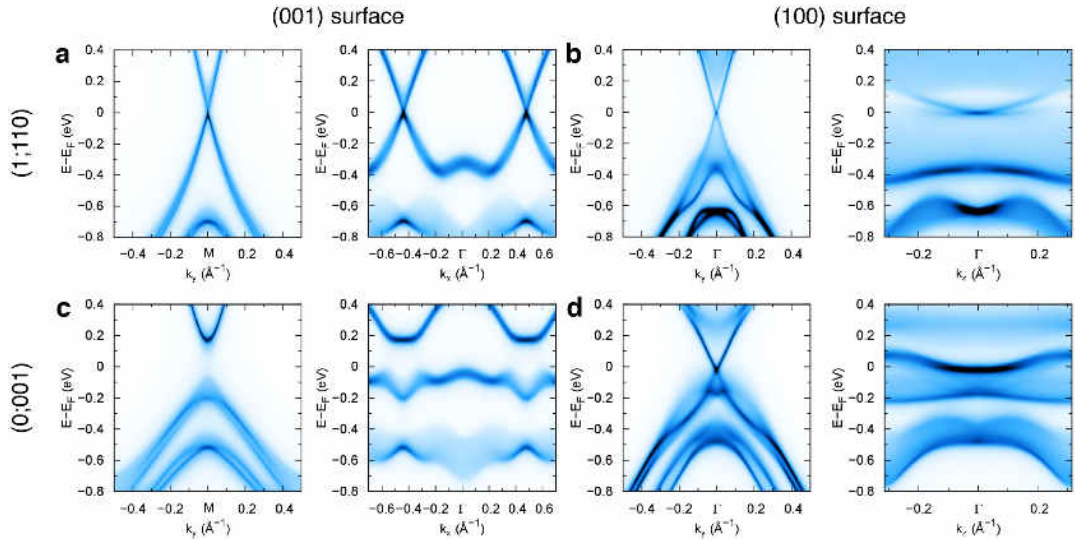

**Supplementary Figure S6. a,b,** Momentum-resolved surface local density of states (LDOS) for the strong topological phase (1;110) at the (001) surface and the (100) surface, respectively. Dispersion along $k_y$ and $k_x$ are shown for the (001) surface while $k_y$ and $k_z$ direction are shown in panel (**b**). See Supplementary Fig. S5a for the definition of $k_x$, $k_y$ and $k_z$. Panel (**a**) reproduces Fig. 2e,g of the main text. **c,d,** Momentum-resolved surface LDOS for the weak topological phase (0;001) at the (001) surface and the (100) surface, respectively.

**Supplementary Table S2.** Correspondence between the TRIM points in the 2D Brillouin zones (BZs) of the (100) and (001) surfaces and those in the bulk BZ. The resulting values of $\pi_{\bar{\Lambda}}$ for the (0;001) weak topological phase are given.

| (100) surface | | | (001) surface | | |
|---|---|---|---|---|---|
| 2D TRIM point | 3D TRIM point | $\pi_{\bar{\Lambda}}$ | 2D TRIM point | 3D TRIM point | $\pi_{\bar{\Lambda}}$ |
| $\bar{\Gamma}$ | Γ, Y | −1 | $\bar{\Gamma}$ | Γ, A | 1 |
| $\bar{Z}$ | A, M | −1 | $\bar{M}$ | M, Y | 1 |
| $\bar{Y}$ | V, V' | 1 | $\bar{L}$ | L, V | 1 |
| $\bar{M}$ | L, L' | 1 | | | |

## 3. Resistivity measurements

Resistivity as a function of temperature, $\rho(T)$, was measured along the longer axis of $\beta$-$Bi_4I_4$ single crystals in a conventional four point configuration. For electrical contacts four Cr/Au stripes, 10/50 nm thick respectively, were evaporated on the sample surface and 25 µm gold wires were glued on them by silver paste. The samples were kept in argon atmosphere prior to measurements. Supplementary Figure S7 presents the resistivity measured in the 4−300 K temperature range. The results are close to that reported in Ref. 22. The room temperature resistivity is $2.5 \times 10^{-3}$ $\Omega$cm. At high temperature the sample displays a semiconducting behavior ($d\rho/dT < 0$). The fitting of $\rho$ above 170 K by $\rho_0 \exp(\Delta/k_B T)$ (Supplementary Fig. S7, inset) gives an activation energy value of $\Delta$ = 36 meV. With lowering temperature a maximum appears at about 150 K and resistivity starts decreasing slightly down to 90 K, and then it increases again below this temperature. It must be noted that a variability was observed in absolute value and in the minimum and maximum position of $\rho(T)$ between different samples. This is probably due to the inhomogeneity in iodine content between the samples.

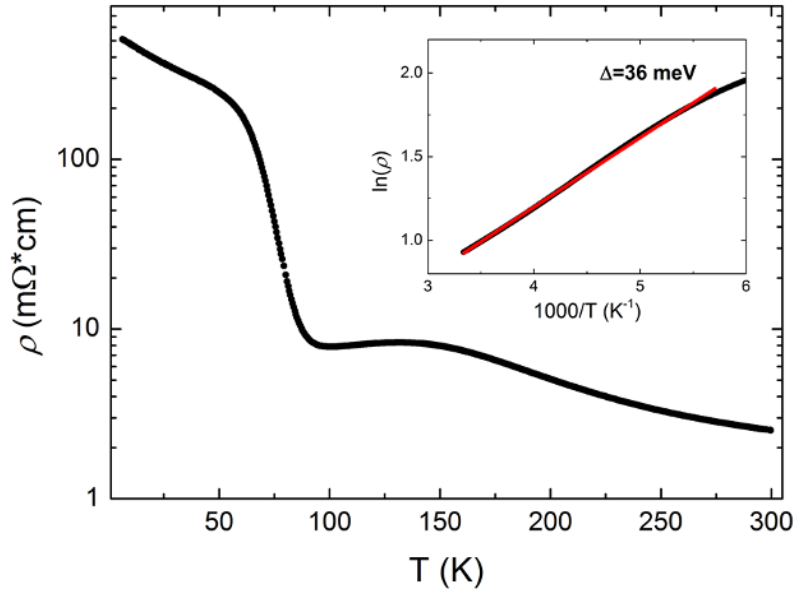

**Supplementary Figure S7.** Resistivity of $\beta$-$Bi_4I_4$ single crystal as a function of temperature. The inset shows the Arrhenius plot at high temperatures serving to extract the activation energy ($\Delta$). The value obtained from the fit (red line) is 36 meV.

# 4. Angle-resolved photoemission spectroscopy (ARPES) measurements

ARPES measurements were performed at beamline 4.0.3 (MERLIN) of the Advanced Light Source, Berkeley. The experimental geometry is sketched in Supplementary Fig. S8. We used $p$ (horizontal) polarization, which provides higher flux. By symmetry arguments, it enables detection of the $p_x$ and $p_z$ orbitals, which constitute the dominant characters close to the Fermi level at the $\bar{M}$ point. Conversely, $s$ polarized light would be sensitive only to $p_y$ orbitals, oriented along the chains.

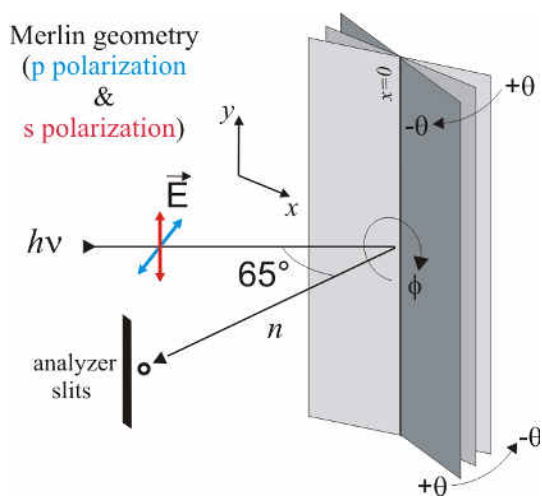

**Supplementary Figure S8.** Experimental geometry used at the Merlin beamline.

Supplementary Figs. S9a–e show a close-up of the $\Lambda$-shaped band, measured at five different photon energies. The crossing point is very close to the Fermi level, so the two branches of the conduction band are not clearly distinguishable due to experimental limitations. Note that the Fermi vector is ~0.01 Å$^{-1}$ and the momentum range is less than 0.2 Å$^{-1}$.

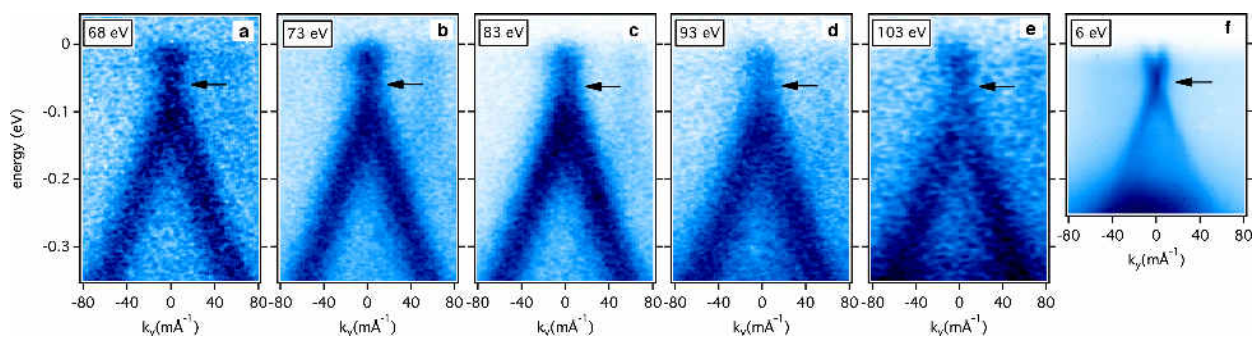

**Supplementary Figure S9.** **a–e,** Close-up of the $\Lambda$-shaped band measured at the Merlin beamline at a temperature of 45 K and different photon energies and, **f**, with a 6 eV laser at a temperature of 17 K (reproduces Fig. 3f of the main text). The $\bar{M}$ point measured here is at $k$ = (0.45, 0) Å$^{-1}$ as in Fig. 3 of the main text, as opposed to $k$ = (0, 1.4) Å$^{-1}$ in Supplementary Fig. S11 below. Arrows indicate the energy of the narrowest MDC. In each image (**a–e**), the sample has been aligned at the $\bar{M}$ point within 0.05°.

In Supplementary Fig. S9f, as well as in Fig. 3f of the main text, the same region is measured with a 6 eV laser as excitation source that, owing to the much smaller kinetic energy of the photoelectrons, enables an improved momentum resolution. The dispersion above the crossing point is now clearly visible. Arrows point to the energy where the momentum distribution curve (MDC) is narrowest, which is associated to the position of the crossing. The finite intensity at the crossing point has contributions from the tails of the bulk VB and CB states, but mostly derives from the topological surface states filling the gap due to the surface sensitivity of the technique, as discussed in the main text. The location of the crossing point energy does not vary both within the synchrotron data and in comparison with the laser data, where the crossing point is more readily discernible, and can always be assigned to a binding energy of 60 ± 5 meV. Depending on the cleave, the presence of defects or adsorbates from the vacuum chamber can shift slightly this point from one sample to another (by at most ± 20 meV). This is not unusual in materials with a very low density of states at the Fermi level, but within the same sample the energy of the band crossing shows no dispersion.

The apparent similarity of the images in Supplementary Fig. S9 points to a negligible dispersion of the electronic states in direction normal to the (001) surface. As discussed in the main text, it is not possible to unambiguously discern non-dispersing surface states from the valence band states that constitute the crossing observed at the $\overline{M}$ point. This is due both to the small magnitude of the band gap and the very weak $k_z$ dispersion of the bulk valence band (37 meV and 15 meV, respectively, according to our $GW$ calculations). We further investigated the $k_z$ dispersion of pure bulk states at approximately 0.35 eV and 0.95 eV binding energy along the Γ−A high-symmetry line. Supplementary Figs. S10a–b follows these two bands as a function of $k_z$ after treatment by the 1D curvature method applied along the energy direction[43]. Their total dispersion is as small as 10−15 meV. Knowing the Γ−A distance of 0.31 Å$^{-1}$ and the location of the band maxima at Γ predicted by theory, we obtained the inner potential value of 16 eV. The weak $k_z$ dispersion of bulk states is a consequence of the quasi-1D structure of the investigated material, as this direction corresponds to interchain coupling.

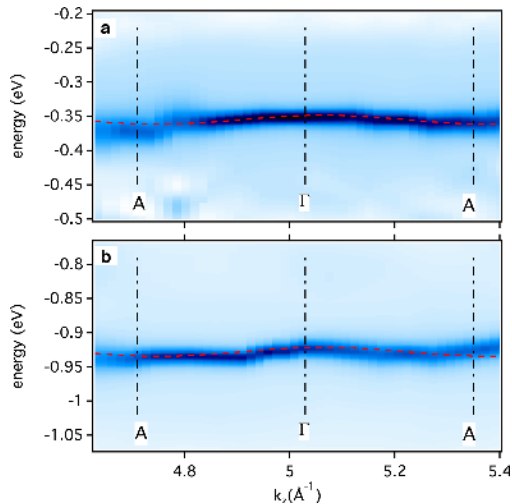

**Supplementary Figure S10. a−b**, Curvature method applied to two valence bands measured along the A−Γ−A line. The dashed sinusoids are guides to the eye and have the period of the crystal Brillouin zone.

The deposition of alkali atoms on the sample surface is suitable for attaining a more complete view of the valence and conduction bands, since in stoichiometric $\beta$-Bi$_4$I$_4$ the Fermi level is very close to the crossing point. We have performed a set of measurements for increasing coverage of K atoms at the surface, as shown in Supplementary Fig. S11. Saturation was not reached, indicating that even the maximum coverage (Supplementary Fig. S11d) is less than one monolayer. This was also confirmed by the presence of a single feature in photoemission from the K 3$p$ level (not shown). Alkali metals, acting as electron donors, enable a data analysis as the one in Supplementary Fig. S12 (see below). However, it should be taken into account that the K deposition leads to a broadening of the linewidth, due to i) an increased disorder at the surface, and ii) a small, but non negligible desorption of the K atoms under the beam, which limits the acquisition time for a given K coverage.

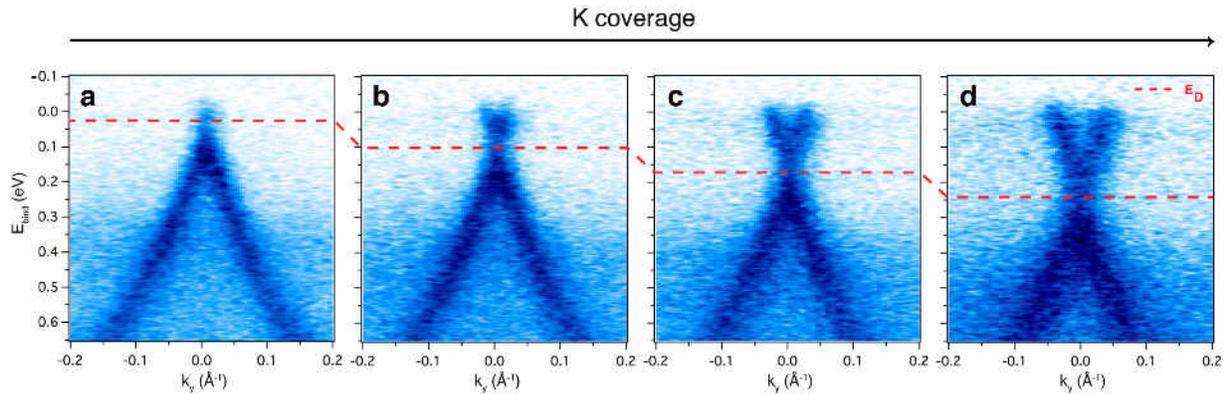

**Supplementary Figure S11.** ARPES spectra acquired during the controlled deposition of potassium atoms from an effusion cell onto the (001) surface of $\beta$-Bi$_4$I$_4$ ($h\nu$ = 85 eV). Potassium was adsorbed at a temperature of 85 K and a partial pressure of $5 \times 10^{-10}$ mbar in consecutive steps in between each ARPES measurement. **a–d,** Photoemission intensity at the $\bar{\text{M}}$ point for different K coverages as a function of binding energy $E_{\text{bind}}$ and momentum $k_y$ parallel to the surface. The shift of the crossing point towards higher binding energies is indicated by the red dashed line.

Supplementary Figure S12a shows an ARPES intensity map measured after K deposition where the crossing is located at −0.3 eV, i.e. approximately 0.25 eV below the value observed for the pristine surface. As noted both here and in the main text, the intensity of the surface states is expected to be dominant close to the band crossing. This makes any assignment of the ARPES signal to bulk or surface states unreliable in proximity of the gap. In the attempt of estimating the bulk gap size from our ARPES data we limit our analysis to energies at least 0.1 eV away from the crossing. Supplementary Figure S12b shows the quasiparticle energies of the valence and conduction bands extracted from fits of the momentum distribution curves (MDCs). The polynomial functions are guides to the eye. Even though such analysis cannot be extended to the band extrema, the range where the fit is reliable is largely sufficient to conclude that the bulk valence and conduction bands cannot be approximated by a single curve. An extrapolation to $k_y = 0$ of the dashed curves in Supplementary Fig. S12b yields a gap magnitude of *ca.* 0.050 eV. This can be considered as a lower limit of the gap at the $k_z$ value corresponding to the photon energy of this measurement ($h\nu$ = 73 eV), since such extrapolated bands would result in an unphysical non-singular derivative at $\bar{\text{M}}$. However, this conclusion should be taken with a pinch of salt since an apparent offset of the bands close to the Dirac point can also result from self-energy effects, as in the known case of plasmarons in graphene[44]. Finally, it is important to note that the observed gap cannot be a spurious effect, given the high estimated accuracy (0.05°) of the sample alignment.

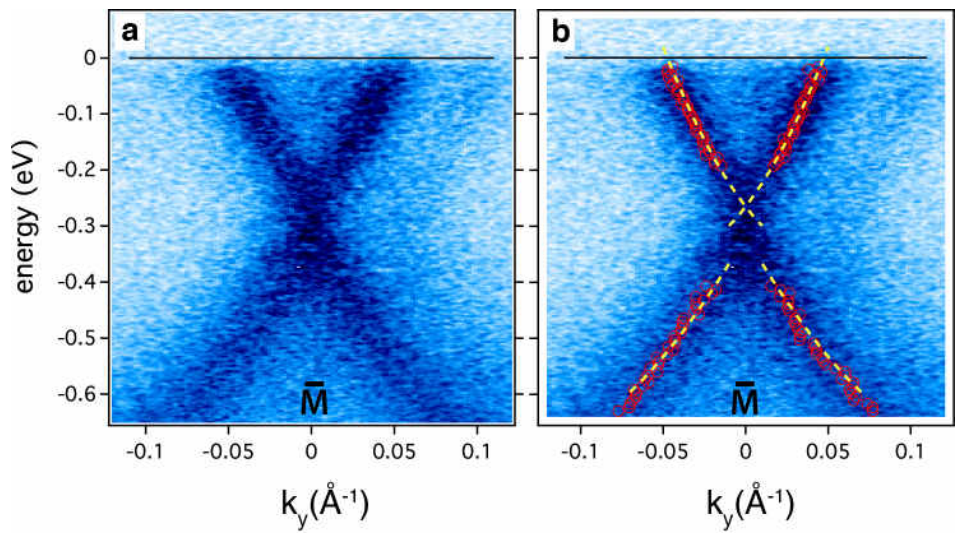

**Supplementary Figure S12. a,** Band dispersion in vicinity of the $\overline{M}$ point [$k = (0, 1.4)$ Å$^{-1}$] after potassium evaporation. **b,** Peak positions obtained from MDC fits are superimposed on the image plot for the ranges of VB and CB where the crossing bands are clearly distinguishable. The dashed lines are polynomial curves attempting to reproduce the dispersion of the bulk bands as determined by the MDC fits.

## 5. Supplementary references